\documentclass{iopconfser}
\usepackage{comment}
\usepackage{amsmath}
\usepackage{hyperref}

\newcommand{\ali}[1]{\begin{align} #1 \end{align}}
\newcommand{\p}{\partial}
\newcommand{\vev}[1]{\langle #1 \rangle}  
\newcommand{\mn}{{\mu\nu}}
\newcommand{\ra}{\rightarrow}

\DeclareMathOperator{\sech}{sech}

\begin{document}

\title{Radial canonical $\Lambda<0$ gravity} 

\author{Nele Callebaut and Blanca Hergueta}

\affil{Institute for Theoretical Physics, University of Cologne, Zülpicher Str.~77a, \\50937 Cologne, Germany}

\email{nele.callebaut@thp.uni-koeln.de, hergueta@thp.uni-koeln.de}

\begin{abstract}
We apply an ADM deparametrization strategy to radial canonical $\Lambda < 0$ gravity in three dimensions. 
It gives rise to a concise notation for previous holographic interpretations  in terms of an identified radial `volume time' and `true' ADM degrees of freedom.   
We further discuss York time and conformal boundary conditions in this context, and construct a BTZ wavepacket solution to the radial WdW equation.  
\end{abstract}

\tableofcontents

\section{Introduction} 

This proceeding is based on the paper \cite{Blacker:2024rje}, which is joint work with Matthew Blacker and Sirui Ning. 
In that work, we are interested in studying the canonical, or Wheeler-DeWitt (WdW), approach to quantizing gravity, applied to gravity with negative cosmological constant $\Lambda < 0$.  
The goal is to make contact to the holographic approach to quantum gravity, which is best understood for the negative sign of the cosmological constant in the form of the AdS/CFT correspondence.  The original, WdW approach quickly runs into problems (such as the treatment of time, defining an inner product and constructing a Hilbert space). 
AdS/CFT on the other hand operates on the (well-substantiated) assumption that indeed $\Lambda < 0$ quantum gravity has a holographic description, and as such offers the answers to the problems of quantum gravity: these answers are hidden in an intricate encoding in conformal field theory (CFT) data. The holographic approach consists of deciphering the dictionary from conformal field theory to gravitational language.  
It is natural to wonder how to connect the original problems to the AdS/CFT `answers'. We take this as a motivation to  
study $\Lambda < 0$ gravity, going forward referred to as AdS-gravity, from a WdW perspective. A non-exhaustive list of references that have taken this approach 
is \cite{Yang:2018gdb,Harlow:2018tqv,Maldacena:2019cbz,PapadoulakiChowdhury:2021nxw,Witten:2022xxp,WallAraujo-Regado:2022gvw,Hartnoll:2022snh}. In particular, \cite{WallAraujo-Regado:2022gvw} makes claims towards WdW formulations of AdS/CFT. For this, they make use of ``$T\bar T$'', a particular deformation of CFT that provides a bridge between AdS/CFT and WdW language. 
Namely, it was shown in \cite{McGough:2016lol,Freidel:2008sh} that the $T\bar T$ path integral can be holographically identified  
with the gravitational wavefunctional that solves the radial WdW equation in AdS$_3$-gravity.

We revisit the radial WdW equation in AdS$_3$-gravity by applying an ADM deparametrization strategy (Section \ref{sectionpara}-\ref{sectiondepara}). It involves identifying volume time as the natural radial evolution direction for a Schr\"odinger equation interpretation of the WdW equation, similar as in cosmological discussions.   
This leads to a concise notation for previous holographic interpretations,  discussed in Section \ref{sectionholo}, and shows that the radial ADM formalism is able to identify CFT 
data (conformal boundary metric and momenta) as `true' degrees of freedom of AdS-gravity. In Section \ref{sectionLaplace}, we 
discuss a Laplace transform between the volume time 
and a York time description of AdS$_3$-gravity. 
This corresponds to 
considering different choices of boundary conditions to the gravitational problem.     
A second way in which we study radial canonical AdS$_3$-gravity is through the construction of the classical BTZ \cite{Banados:1992wn} solution from the Hamilton-Jacobi equation, and a corresponding semi-classical wavepacket solution from the radial WdW equation (Section \ref{sectionBTZ}-\ref{sectionwavepacket}).

\section{Parametrization in mechanics}  \label{sectionpara} 

To review the idea of ADM deparametrization, we start with a simple mechanical system with $M$ degrees of freedom, described by an action 
\ali{
	S = \int dt \, \left(p_i \p_t q_i - H(p_i, q_i,t) \right), \qquad (i = 1 \cdots M).  \label{Sone}
}
By interpreting time and Hamiltonian as an extra conjugate pair $(t, -H) \equiv (q_{M+1}(\tau), p_{M+1}(\tau))$, in addition to the $M$ conjugate pairs of coordinates and momenta $(q_i, p_i)$, we can bring $S$ in a form parametrized by the arbitrary parameter $\tau$, 
\ali{
	S = \int d\tau \, \left. \left(p_j \p_\tau q_j \right)\right|_{p_{M+1} = -H} , \qquad (j = 1 \cdots M+1).  \label{Sonebis}
} 
We will distinguish between the original set of conjugate pairs and the extended one by using labels $i$ and $j$ running over respectively $M$ or $M+1$ values. 
The condition identifying $p_{M+1}$ with $-H$ can be enforced by a Lagrange multiplier $N(\tau)$, 
\ali{
	S = \int d\tau \, \left(p_j \p_\tau q_j - N \mathcal H \right) , \qquad (j = 1 \cdots M+1) \label{Stwo}, 
} 
with variation with respect to $N$ imposing the constraint $\mathcal H(q_i, p_i,q_{M+1},p_{M+1}) = 0$ with solution $p_{M+1} = -H(q_i, p_i,q_{M+1})$. 

The first description \eqref{Sone}  
comes in a `deparametrized' form of the action, and the second \eqref{Stwo} in a `parametrized' form. The latter has more degrees of freedom, but also more symmetry, in the form of reparametrization invariance under  $\tau \ra \tau'(\tau)$. 
Reversely, to go from the parametrized to deparametrized description,  \eqref{Stwo} to \eqref{Sone}, requires a two-step process: first substituting the constraint solution imposed by varying $N$, and subsequently imposing what is called a ``coordinate condition'' in the original ADM paper \cite{Arnowitt:1962hi}. The latter step is a gauge choice or identification of a `preferred' time, $q_{M+1}(\tau) = \tau$ combined with calling $q_{M+1} \equiv t$.

Schematically, the variation of the action \eqref{Sonebis} takes the form 
$\delta S = \int (\text{EOM}) \delta q_j  + p \delta q|_{b} - H \delta t|_{b}$,
such that the on-shell variation \cite{Banados:2016zim} $\delta S_{cl}$ only  contains boundary contributions, from which one can read off  
\ali{
	p_{cl}(\tau_b) = \frac{\delta S_{cl}}{\delta q_{b}}, \qquad H_{cl}(\tau_b) = -\frac{\delta S_{cl}}{\delta t_{b}} .  \label{eq4}
}
Here, the subscript `cl' refers to evaluation on a classical solution $q_j(\tau)$  
and the subscript `b' to evaluation in the boundary values 
of the integration variable $\tau = \tau_{b}$, i.e.~$q_{b} \equiv q(\tau_{b})$ and $t_b \equiv t(\tau_b)$.    
This leads to the 
Hamilton-Jacobi equation 
\ali{
	\frac{\delta S_{cl}}{\delta t} = -H(q,\frac{\delta S_{cl}}{\delta q}, t)  \label{HJmech} 
} 
where now explicit subscripts (except for $S_{cl}$) have been dropped for readability. 

\section{Deparametrization in canonical gravity}  \label{sectiondepara} 

In the Hamiltonian formulation of General Relativity, in $(3+1)$-dimensional $\Lambda = 0$ gravity originally, one makes use of the ADM parametrization of the metric 
\ali{
	ds^2 = -N^2 d\tau^2 + \gamma_{ij}(dx^i + N^i d\tau) (dx^j + N^j d\tau)   
}
with $N(\tau, x)$ the lapse, $N^i(\tau, x)$ the shift, and $\gamma_{ij}(\tau,x)$ the intrinsic metric on spatial slices $\Sigma$ that foliate the spacetime. In terms of these fields, the gravitational action (up to boundary contributions \cite{Brown:1992br}) takes the form 
\ali{
	S = \int d^3 x d\tau \, \left( \pi^{ij} \p_\tau \gamma_{ij} - N \mathcal H - N^i \mathcal H_i \right).  \label{Sgravpara}
}
It can thus be recognized as being of a deparametrized form such as \eqref{Stwo}.   
Here, $(\gamma_{ij}, \pi^{ij})$ take the role of conjugate pairs $(q_j, p_j)$ of the previous section, the momenta $\pi^{ij}$ defined from the Lagrangian density $\mathcal L$ as 
\ali{
	\pi^{ij} = \frac{\p \mathcal L}{\p \, \p_\tau \gamma_{ij}}.  
}
The lapse and shift impose respectively the Hamiltonian constraint 
\ali{
	\mathcal H = 0 
}
and momentum constraint $\mathcal H_i = 0$, with $\mathcal H$ quadratic in the momenta. Upon canonical quantization, the momenta are promoted to variational derivatives with respect to the 3-metric, $\pi^{ij} \ra -i \delta/\delta \gamma_{ij}$, as they act on a gravitational wavefunctional $\psi$ in the `coordinate representation' $\psi(\gamma_{ij}) = \int_{g_{ij}(\Sigma) = \gamma_{ij}} \mathcal Dg \exp{ \{ i S[g] \} }$. The Hamiltonian constraint then becomes a condition on allowed $\psi$ in the form of a differential equation called the WdW equation 
\ali{
	{\mathcal H} \, \psi = 0. 
}
There are normal ordering ambiguities 
concerning the action of the momenta as derivative operators. In the semi-classical approximation, however, where $\psi$ is dominated by its saddle point, 
the WdW equation reduces back to the gravitational Hamilton-Jacobi equation, 
which is obtained from the Hamiltonian constraint by replacing $\pi^{ij}$ by their classical value $\delta S_{cl}/\delta \gamma_{ij}$. 

To bring \eqref{Sgravpara} in a deparametrized form, in analogy to the discussion in the mechanical system, one requires a two-step process involving both substituting the solution to the constraints and a coordinate condition \cite{Arnowitt:1962hi}. Discussions of such a procedure in $(2+1)$-dimensional $\Lambda = 0$ gravity can be found in \cite{Martinec:1984fs,Moncrief:1989dx}.

\subsection{Applied to radial canonical $\Lambda < 0$ gravity}  

Applying the previous discussion to AdS$_3$-gravity with $\Lambda = -1/l^2$ (for AdS radius $l$), we work with a \emph{radial} ADM parametrization 
\ali{
	ds^2 = N^2 dr^2 + \gamma_{ij}(dx^i + N^i dr) (dx^j + N^j dr)  \label{radialADM}
}
of the metric $ds^2 = g_\mn dx^\mu dx^\nu$ (with Latin indices taking two and Greek indices three values). The $x^i$ coordinates consist of one space and one time coordinate. 
The gravitational action 
\ali{
	S = \frac{1}{2\kappa} \int d^3 x \sqrt{-g} \left( R_{(3)} + \frac{2}{l^2} \right) - \frac{1}{\kappa} \int d^2 x \sqrt{-\gamma} K - \frac{1}{\kappa l} \int d^2 x \sqrt{-\gamma}  \label{Sgravstart}
}
including the counterterm \cite{Balasubramanian:1999re} then again takes the parametrized form 
\ali{
	S = \int d^2 x dr \, \left( \pi^{ij} \p_r \gamma_{ij} - N \mathcal H - N^i \mathcal H_i \right)   \label{Sgrav}
}
with momenta $\pi^{ij} = -\frac{1}{2\kappa} \sqrt{-\gamma} (K \gamma^{ij} - K^{ij} + \frac{1}{l} \gamma^{ij})$, Hamiltonian density  
\ali{
	\mathcal H = -\frac{\sqrt{-\gamma}}{2\kappa} R_{(2)} - \frac{2\kappa}{\sqrt{-\gamma}} (\pi^{ij} \pi_{ij} - \pi^2 ) + \frac{2}{l} \pi \, , \label{Hgrav}
}
and momentum density $\mathcal H_i = -2 \nabla_j^{(2)} \pi_i^j$. 
For later reference, the gravitational Hamilton-Jacobi equation (from  $\pi^{ij} \ra \delta S_{cl}/\delta \gamma_{ij}$ in $\mathcal H = 0$) is thus 
\ali{
	-\frac{l}{2\kappa} R_{(2)} - \frac{2\kappa l}{\gamma} \left(\frac{\delta S_{cl}}{\delta \gamma_{ij}} \frac{\delta S_{cl}}{\delta \gamma_{kl}} \gamma_{ki} \gamma_{jl} - \left(\frac{\delta S_{cl}}{\delta \gamma_{ij}} \gamma_{ij}\right)^2 
	\right) +  \frac{2}{\sqrt{-\gamma}} \frac{\delta S_{cl}}{\delta \gamma_{ij}} \gamma_{ij}
	= 0 , 
}
which should be read as an equation for the on-shell action's dependence on boundary data $\gamma_{ij}$.

To proceed, we first make a symmetric, `mini-superspace' ansatz for the metric, in which all metric fields only depend on the radial coordinate: $N = N(r)$, $N^i = N^i(r)$, and $\gamma_{ij} = \gamma_{ij}(r)$. 
This implies the 
hypersurfaces $\Sigma$ are non-intersecting, 
and the momentum constraint is automatically satisfied $\mathcal H_i \equiv 0$. Then we perform the exercise of  
applying an ADM deparametrization procedure to the remaining system, i.e.~the action \eqref{Sgrav} (with $\mathcal H_i \equiv 0$) and corresponding Hamiltonian constraint $\mathcal H = 0$ in \eqref{Hgrav}. This results in a deparametrized action 
\ali{
	S = \int dv \, d^2 x \left( \tilde \pi^{ij} \p_v \tilde \gamma_{ij} - \mathcal H_{ADM} \right) 
}
and corresponding 
rewritten Hamiltonian constraint  
\ali{
	\pi_v = - \mathcal H_{ADM}, \label{gravpiv}
} 
where 
\ali{
	\mathcal H_{ADM}(\tilde \gamma_{ij}, \tilde \pi^{ij}, v) \equiv \frac{1}{\kappa l} - \frac{1}{\kappa l} \sqrt{1 + 2 \kappa^2 l^2 \frac{\tilde \pi^{ij} \tilde \pi_{ij}}{v^2} + \frac{l^2 R_{(2)}}{2}}.  
} 
The new coordinates that appear are the volume density $v$ of the slices, its canonical partner $\pi_v$, which is known as York time in $\Lambda = 0$ gravity literature \cite{York1:1972sj}, the metric $\tilde \gamma_{ij}$ with unit determinant $\sqrt{-\tilde \gamma} = 1$, and finally its canonical partner, the traceless momentum $\tilde \pi^{ij}$. They are defined as 
\ali{
	v = \sqrt{-\gamma}, \quad \pi_v = \frac{\pi}{\sqrt{-\gamma}}, \quad \tilde \gamma_{ij}= \frac{\gamma_{ij}}{\sqrt{-\gamma}}, \quad \tilde \pi^{ij} = \sqrt{-\gamma} (\pi^{ij} - \frac{1}{2} \gamma^{ij} \pi)   \label{vcoord}
} 
with $\pi^i_i \equiv \pi$. 
As such, the original canonical pairs are decomposed 
\ali{
	(\gamma_{ij}, \pi^{ij}) \ra (v, \pi_v, \tilde \gamma_{ij}, \tilde \pi^{ij}) 
}
in what are called the `Dirac variables' 
in \cite{Regge:1974zd} (satisfying indeed $\tilde \gamma = 1$, $\tilde \pi = 0$ and establishing an orthogonal decomposition \cite{Blacker:2024rje,York2:1973ia}). In the mechanics language of Section \ref{sectionpara}, we have arrived at a deparametrized action by substituting a constraint \eqref{gravpiv} in the form $p_{M+1} = -H$, and identifying a preferred radial time\footnote{We set $l=1$ here, and bring it back in Section \ref{sectionholo}. 
} $r=-v$, a corresponding true or `ADM' radial Hamiltonian density $\mathcal H_{ADM} = -\pi_v$, as well as true or `ADM' degrees of freedom $(\tilde \gamma_{ij}, \tilde \pi^{ij})$. The equivalence between the mechanics and the radial gravity discussion is summarized in Table \ref{table:1}. 
\begin{table}[h!]
\begin{center}
	\begin{tabular}{ |c|c|c| } 
		\hline
		& Mechanics & Radial canonical gravity \\
		\hline
		`time' parameter & $\tau$ & $r$ \\ 
		degrees of freedom & $(p_j,q_j)$ \quad {\scriptsize $(j = 1, ..., M+1)$},  & $(\pi^{ij},\gamma_{ij})$ \\  
		preferred `time'& $q_{M+1}(\tau) = \tau \equiv t$ & $-\sqrt{-\gamma(r)} = r \equiv -v$ \\  
		true Hamiltonian & $p_{M+1}\equiv -H(p_i,q_i,q_{M+1})$ & $\pi_v \equiv - \mathcal H_{ADM}(\tilde \pi^{ij},\tilde \gamma_{ij},v)$ \\  
		true degrees of freedom & $(p_i,q_i)$ \quad \quad {\scriptsize $(i = 1, ..., M)$} & $(\tilde \pi^{ij},\tilde \gamma_{ij})$ \\  
		\hline
	\end{tabular} 
\end{center}
\caption{Comparison between Hamiltonian description of mechanics and AdS$_3$-gravity (for $l=1$).} 
\label{table:1}
\end{table} 

We will refer to our choice of radial time as \emph{volume time}. It is familiar from de Sitter (dS) cosmology \cite{Hawking:1983hn} where actual time corresponds to expansion of the universe. In \emph{Anti} de Sitter (AdS), it then naturally defines a \emph{radial} time. It is a well-defined direction of evolution 
until the maximal volume slice is reached. 

Upon quantization, the condition \eqref{gravpiv} becomes a volume time Schr\"odinger equation 
\ali{
	-i \frac{\delta}{\delta v} \psi = H_{ADM} \, \psi   \label{volschrod}
}
for ADM Hamiltonian $H_{ADM} = \int d^2 x \, \mathcal H_{ADM}$ and $\psi$ the gravitational wavefunctional for Dirichlet boundary conditions imposed at the slice $\Sigma$, i.e.~$\psi(\gamma_{ij}) = \int_{g_{ij}(\Sigma) = \gamma_{ij}} \mathcal Dg \exp{ \{ i S[g] \} }$. This provides us with a Schr\"odinger equation interpretation of the WdW equation, analogous to that in 
\cite{Kuchar:1971xm}. 
In the semi-classical 
regime, it reduces to a gravitational Hamilton-Jacobi equation of the form \eqref{HJmech}, 
\ali{
	\frac{\delta S_{cl}}{\delta v} = - H_{ADM}(\tilde \gamma_{ij}, \frac{\delta S_{cl}}{\delta \tilde \gamma_{ij}},v).  
}
Indeed, the on-shell variation of the action $\delta S_{cl} = \int d^2 x \, (\pi^{ij} \delta \gamma_{ij})$ splits into 
\ali{
	\delta S_{cl}  
	= \int d^2 x \, (\tilde \pi^{ij} \delta \tilde \gamma_{ij} + \pi_v \delta v)    \label{deltaScl}
}
thanks to the properties $\tilde \gamma = 1$ and $\tilde \pi = 0$ of the ADM degrees of freedom, again mimicking the mechanics discussion for deriving \eqref{HJmech}. (Here we suppress again extra `cl' subscripts on the classical momenta for readability, as in \eqref{eq4} to \eqref{HJmech}.)

\section{Holographic interpretation} \label{sectionholo}

In this section we move on to the holographic interpretation of the ADM deparametrization results. 

The true degrees of freedom $\tilde \gamma_{ij}$ represent the  
flat conformal boundary metric of AdS$_3$,  
and the asymptotic ($v \ra \infty$) WdW equation contains Weyl anomaly physics of the dual CFT, namely its behavior under variation of scale or volume. 
Indeed, the small $\kappa$ (or large $v$, near-boundary) 
Hamilton-Jacobi equation 
\eqref{gravpiv} takes the form 
\ali{ 
	\pi_v = \frac{l}{4 \kappa} R_{(2)},
} 
which expresses the Weyl anomaly  
\ali{
	\vev{T^i_i} = \frac{c}{24\pi} R_{(2)} 
}
of the large $c$ holographic CFT. 
The CFT has a central charge $c = 12 \pi l/\kappa$ \cite{Brown:1986nw} and stress tensor expectation value given by the Brown-York stress tensor 
$T^{ij} = 2 \pi^{ij}/\sqrt{-\gamma}$  
for the corresponding classical gravity solution (with $T^i_i = 2 \pi_v$).  
At the level of the WdW equation \eqref{volschrod}, its near-boundary form 
\ali{
	-i \frac{\delta}{\delta v} \psi = \frac{l}{4 \kappa} R_{(2)} \psi 
	}
expresses the conformal Ward identity 
\ali{
	-i \frac{\delta}{\delta v} Z_{CFT} = \frac{c}{48\pi} R_{(2)} Z_{CFT} 
}
of the dual CFT, with path integral $Z_{CFT}(\gamma) = \int \mathcal D \phi \exp{ \{i S_{CFT}[\phi] \}}$.  

The above gives a volume time reformulation of Freidel's formulation of AdS/CFT \cite{Freidel:2008sh} that the asymptotic WdW equation is solved by $\psi = Z_{CFT}$, and of holographic renormalization  \cite{Heemskerk:2010hk,Henningson:1998gx}. 

Removing the near-boundary limit, the full Hamilton-Jacobi equation  
\eqref{Hgrav} 
\ali{
	\pi_v = \frac{l}{4\kappa} R_{(2)} + \frac{\kappa l}{v^2} \tilde \pi^{ij} \tilde \pi_{ij}- \frac{\kappa l}{2} \pi_v^2 
    }
or 
\ali{
	\frac{2}{\sqrt{-\gamma}} \frac{\delta S_{cl}}{\delta \gamma_{ij}} \gamma_{ij} = 	\frac{l}{2\kappa} R_{(2)} + \frac{2\kappa l}{\gamma} \left(\frac{\delta S_{cl}}{\delta \gamma_{ij}} \frac{\delta S_{cl}}{\delta \gamma_{kl}} \gamma_{ki} \gamma_{jl} - \left(\frac{\delta S_{cl}}{\delta \gamma_{ij}} \gamma_{ij}\right)^2 
	\right)  
}
has been 
recognized in \cite{McGough:2016lol} to express the trace flow equation 
\ali{
	\vev{T_i^i} = \frac{c}{24 \pi} R_{(2)} - \lambda \left( \vev{T_{ij}} \vev{T^{ij}} - \vev{T^i_i}^2 \right)   
}
of dual $T\bar T$ theories \cite{Smirnov:2016lqw} with $T\bar T$ coupling $\lambda = \kappa l/2$.  
In \cite{Freidel:2008sh}, a closed expression was given for the $\psi$ solution to the radial WdW equation in vielbein formalism, in the form of 
an integral transform of the CFT partition function. It is this closed form in vielbein notation that has been interpreted as $T\bar T$ \cite{McGough:2016lol,Tolley:2019nmm}.

\section{Laplace transform between Dirichlet and conformal boundary condition problem} \label{sectionLaplace}

For this section we switch to Euclidean signature. This leads to some sign changes, including the action $S$ in \eqref{Sgravstart} to $-S$.   
The Euclidean gravitational wavefunctional 
\ali{
	\psi(v,\tilde \gamma_{ij}) = \int_{g_{ij}(\Sigma) = \gamma_{ij}} \mathcal Dg \exp{ \{ -S[g] \} },  
}
with $\delta S_{cl} = -\int d^2 x \, (\tilde \pi^{ij} \delta \tilde \gamma_{ij} + \pi_v \delta v)$ for the saddle point, can be Laplace transformed \cite{Hawking:1983hn} to   
\ali{
	\phi(\pi_v, \tilde \gamma_{ij})  = \int_0^\infty \mathcal D v \,  e^{-\int d^2 x \, \pi_v v} \psi(v,\tilde \gamma_{ij}).    \label{phiwavef}
} 
It gives a wavefunction $\phi =  \int_{CBC} \mathcal Dg \exp{ \{ - \tilde S[g] \} }$ with a transformed action $\tilde S = S + \int d^2 x \, \pi_v v$  whose on-shell variation 
\ali{
	\delta \tilde S_{cl}  
	= \int d^2 x \, (-\tilde \pi^{ij} \delta \tilde \gamma_{ij} + v \delta \pi_v)    \label{deltatildeScl}
}
is 
the suitable one for conformal boundary conditions (CBC) (in three bulk dimensions, the action with a well-defined variational principle for CBC is the Einstein-Hilbert action plus 1/2 times the Gibbons-Hawking-York term, compared to Einstein-Hilbert plus Gibbons-Hawking-York for Dirichlet boundary conditions). Namely, $\delta \tilde S_{cl}$ vanishes for fixed conformal metric $\tilde \gamma_{ij}$ and fixed `York time' $\pi_v$. Equivalently, as $\pi_v = -(K + \frac{2}{l})/(2\kappa)$, we can replace the $\pi_v$-dependence of $\phi$ in \eqref{phiwavef} by $K$-dependence and $\delta \pi_v$ by $\delta K$ in \eqref{deltatildeScl}. This makes clear that CBC fix the conformal metric and the trace of extrinsic curvature at 
$\Sigma$. These boundary conditions are discussed in \cite{Witten:2022xxp} to lead to well-defined wavefunctions $\phi$ in perturbation theory. 

The corresponding Hamilton-Jacobi equation for $\tilde S_{cl}$ will contain variations $\delta \tilde S_{cl}/\delta \tilde \gamma_{ij}$ and $\delta \tilde S_{cl}/\delta \pi_v$, and the corresponding WdW equation for $\phi$, similarly as the discussion in Section \ref{sectiondepara}, allows an interpretation as a Schr\"odinger-like (in Euclidean signature) evolution equation in \emph{York time} $\pi_v$ or $K$. 

Where in statistical physics the Laplace transform takes us between different ensemble descriptions, 
here the Laplace transform takes us from a Dirichlet problem, with volume time $v$ as radial time, to a conformal boundary condition problem, with York time $K$ as radial time.

\section{BTZ solution from Hamilton-Jacobi} \label{sectionBTZ}

In this section we discuss the classical, non-rotating BTZ solution of AdS$_3$-gravity by solving the gravitational Hamilton-Jacobi equation, and next construct a corresponding wavepacket solution to the WdW equation. In this and the next section, we follow closely \cite{Hartnoll:2022snh}, applied to one lower dimension.

The starting point is the mini-superspace (all metric components functions of $r$ only), radial ADM ansatz of \eqref{radialADM}, with $N^i = 0$  
and in the volume coordinates introduced in \eqref{vcoord}, 
\ali{
	ds^2 = N^2 dr^2 + v \tilde \gamma_{ij} dx^i dx^j = N^2 dr^2 + v \tilde \gamma_{tt} dt^2 + v \tilde \gamma_{tt}^{-1} dt^2. 
}
In the second equality we made use of the fact that $\sqrt{-\tilde \gamma}=1$ (and have set $\gamma_{t\varphi} = 0$ to restrict to non-rotating solutions for simplicity). 
Working with this ansatz, we would obtain the outside-horizon region of BTZ (where $r$ is a spacelike coordinate). To continue constructing instead the inside-horizon region (where $r$ is a timelike coordinate), we work with the ansatz 
\ali{
	ds^2 = -N^2 dr^2 + v e^k \frac{dt^2}{(\Delta t)^2} + v e^{-k} \frac{d\varphi^2}{(\Delta \varphi)^2}.  \label{ansatzvk}
}
Aside from the sign of the $dr^2$ term, there are two more changes. One is using the notation $e^k$ for $\tilde \gamma_{tt}$. 
The second is that 
$v$ is now the actual volume (not volume density) 
of the radial slices $\Sigma$, with dimension length squared.  
This is realized 
by working with rescaled coordinates $t/\Delta t$ and $\varphi/\Delta \varphi$.

With this ansatz, the 
Hamilton-Jacobi equation takes the form 
\ali{
	-\left( \frac{\p S_{cl}}{\p k} \right)^2 + v^2 \left( \frac{\p S_{cl}}{\p v} \right)^2 + \frac{v^2}{\kappa^2 l^2} = 0,   
}
where $S_{cl}$, to be complete, is now referring to the gravitational action excluding the counterterm (in contrast to the notation in previous sections).  
It is solved by 
\ali{ 
	S_{cl} = \frac{v}{\kappa l} \sinh(k + k_0)
	}
with integration constant $k_0$. The Hamilton-Jacobi procedure further imposes $\p S_{cl}/\p k_0 = \epsilon_0$ for $\epsilon_0$ another constant. This can be solved for the metric components in terms of the integration constants, 
\ali{
	v = \kappa l \epsilon_0 \sech(k+k_0) , \label{clsol}
}
and one obtains indeed the BTZ metric 
\ali{
	ds^2 = \frac{1}{z^2} \left( \frac{dz^2}{f(z)} - f(z) d\tilde t^2 + d\varphi^2 \right) 
}
with $z = 1/r$, $\tilde t = t e^{-k_0}$ and $f(z) = 1 - 2\kappa \epsilon_0 z^2$ (for $l=1$). It represents an AdS$_3$-gravity black hole with mass $\epsilon_0$.

\section{BTZ wavepacket} \label{sectionwavepacket}

Going to the quantum level, the WdW equation in the $(v,k)$-notation of Section \ref{sectionBTZ} is given by 
\ali{
	\p_k^2 \psi - v \p_v (v \p_v \psi) + \frac{v^2}{\kappa^2 l^2} \psi = 0. \label{WdWvk} 
}

The applied normal ordering choice is that of Laplace normal ordering. Let us expand a bit on this. 
In WdW gravity, the $\Sigma$-metric $\gamma_{ij}$ takes the role of configuration space coordinates $q^A$.  
The metric $G_{AB}$ on that configuration space is called the DeWitt metric, and the configuration space itself superspace. 
Working with the ansatz \eqref{ansatzvk}, it is the functions $v(r)$ and $k(r)$ that form the superspace coordinates $q^A$ ($A=1,2$). Their corresponding momenta $p_v$ and $p_k$, or $p_A$ ($A=1,2$), are obtained in the usual way from derivatives of the Lagrangian with respect to $\p_r q^A$. 
The DeWitt metric 
can be read off from the gravitational action in Hamiltonian form \eqref{Sgrav} or from the Hamiltonian density \eqref{Hgrav} (evaluated on the ansatz \eqref{ansatzvk}) from matching to the expected kinetic plus potential term expression 
\ali{
	\mathcal H = \frac{1}{2} G^{AB} p_A p_B + V(q^A).  
}
Namely,  
\ali{
	G_{AB} = \begin{pmatrix}
		-1/v & 0 \\ 0 & v  
	\end{pmatrix}   
    \label{GAB}
}
and the Laplace normal ordering \cite{halliwell_derivation_1988} $(-\frac{1}{2}\nabla^2 + V) \psi = (-\frac{1}{2} \frac{1}{\sqrt{-G}} \p_A (\sqrt{-G} G^{AB} \p_B)  + V ) \psi$ gives \eqref{WdWvk}.

Using $\p_k^2 \psi = -\epsilon^2 \psi$ for a $k$-dependence $\psi \sim e^{i k \epsilon}$, the  WdW equation \eqref{WdWvk} takes the form 
\ali{
	-\hbar^2 v \p_v (v \p_v \psi ) + \frac{v^2}{\kappa^2 l^2} \psi = \epsilon^2 \psi ,  
}
where we reinstalled $\hbar$ just for this line.   
This can be studied in a WKB approximation (small $\hbar$), with 
WKB solution 
\ali{
	\psi_\pm(v,k;\epsilon) &= \frac{e^{i \epsilon k}}{(\epsilon^2 - \frac{v^2}{\kappa^2 l^2})^{1/4}} e^{\pm i \int \frac{dv}{v}\sqrt{\epsilon^2 - \frac{v^2}{\kappa^2 l^2}}} 
} 
or 
\ali{ 
	\psi_\pm(v,k;\epsilon) &= \frac{e^{i \epsilon k}}{(\epsilon^2 - \frac{v^2}{\kappa^2 l^2})^{1/4}} \exp{ \left\{ \pm i \left[ \sqrt{\epsilon^2 - \frac{v^2}{\kappa^2 l^2}} - \epsilon \tanh^{-1} \frac{\sqrt{\epsilon^2 - \frac{v^2}{\kappa^2 l^2}}}{\epsilon} - \frac{\pi}{4} \right] \right\} }. 
}
From this semi-classical solution, one can build a wavepacket solution  
by a Gaussian superposition 
\ali{
	\psi_{wp} = \int \frac{d\epsilon}{2\pi} \alpha(\epsilon) \, \psi_\pm(v,k;\epsilon)  ,  
	\label{wavepacket}
}
using a Gaussian wavepacket
\ali{ 
	\alpha(\epsilon) = \left(\frac{2\sqrt{\pi}}{\Delta}\right)^{1/2} e^{i k_0 \epsilon} e^{-(\epsilon-\epsilon_0)^2/(2\Delta^2)} \label{alphaGaussian}
}
that is centered on $\epsilon = \epsilon_0$ with width $\Delta$, and on $k_0$ with width $1/\Delta$.    

To discuss 
the norm of $\psi_{wv}$, we employ the DeWitt norm \cite{dewitt_quantum_1967} associated with a given choice of time direction $q^0$ in the configuration space, i.e.~superspace. It is given by 
\ali{
	\langle \psi| \psi \rangle^2_{\, q^0} \propto - \frac{i}{2} \int dq^1 \dots dq^n \left[ \psi^* \left( \sqrt{-G} \,G^{0A} \partial_{A} 
	\psi \right) - \text{h.c.} \right], \label{DeWittnorm}
}
for an $(n+1)$-dimensional superspace. Applied to our choice of volume time $q^0 = v$, and using \eqref{GAB}, the norm of our wavepacket solution is given by 
\ali{
	\langle \psi_{wp}| \psi_{wp} \rangle^2_{\, v} \propto - \frac{i}{2} \int dk \left( \psi^*_{wp} v \p_v \psi_{wp} - \psi_{wp} v \p_v \psi^*_{wp} \right) = \pm \int \frac{d\epsilon}{2\pi} |\alpha(\epsilon)|^2 = \pm 1 .  
} 
It is a sensible norm for the choice of the plus sign in the mode $\psi_\pm(v,k;\epsilon)$.  
With this choice, we can proceed to calculate several expectation values in the wavepacket solution, for small $\Delta$ and small (but suppressed)  $\hbar$: 
\ali{
	\langle k \rangle_v &=  -k_0 + \sech^{-1} \frac{v}{\epsilon_0  \kappa  l}- \frac{\Delta^2}{4 \epsilon_0^2}\left(1-\left(\frac{v}{\epsilon_0 \kappa l}\right)^2\right)^{-\frac{3}{2}} + \dots , 
}
\ali{
	\langle k^2 \rangle_v&= \left(-k_0 + \sech^{-1} \frac{v}{\epsilon_0  \kappa  l}\right)^2  
	+ \frac{\Delta^2}{2 \epsilon_0^2} \left(k_0- \sech^{-1}\frac{v}{\epsilon_0  \kappa l}\right)\left(1-\left(\frac{v}{\epsilon_0 \kappa  l}\right)^2\right)^{-\frac{3}{2}} + \frac{\Delta^2}{2  \left(\epsilon_0^2 - \frac{v^2}{\kappa^2 l^2} \right)} - \frac{1}{2\Delta^2} + \dots , \nonumber
}
\ali{
	\langle \pi_k\rangle_v &=  \int \frac{d\epsilon}{2\pi}\epsilon |\alpha(\epsilon)|^2 =  \epsilon_0, \label{pik-v-clock}
}
and 
\ali{
	\langle \pi_k^2\rangle_v 
	&= \int \frac{d\epsilon}{2\pi}\epsilon^2 |\alpha(\epsilon)|^2 = \epsilon_0^2 + \frac{\Delta^2}{2}. \label{pik2-v-clock}
}
As expected, for vanishing $\Delta$, they reduce to the corresponding classical  
solutions, see \eqref{clsol}. Whereas $\vev{k}_v$ and $\vev{k^2}_v$   
depend on $v$, the energy $\vev{\pi_k}_v$ is constant in volume time. (This means it is not the object that can be identified with $T\bar T$ energy holographically.) 
One can check that the uncertainty principle holds,   $\text{var}(k)_v\text{var}(\pi_k)_v = (\langle k^2\rangle-\langle k\rangle^2)(\langle \pi_k^2\rangle-\langle \pi_k\rangle^2)= \frac{1}{4}+ \mathcal{O}(\Delta^2)$, as a check on the calculations.  

It will be interesting to study the BTZ wavepacket more closely from a $T\bar T$ perspective.

\section*{Acknowledgments} 
We thank Matthew Blacker and Sirui Ning for collaboration on \cite{Blacker:2024rje}, on which this proceeding is based. This work has been supported by the Deutsche Forschungsgemeinschaft (DFG, German Research Foundation) within CRC network TR 183 (Project Grant No. 277101999).

\bibliographystyle{vancouver} 
\bibliography{referencesWdWDraftvancouverbisbis}

\end{document}